\documentclass[amsmath,amssymb,aps,prb,twocolumn,floatfix,superscriptaddress]{revtex4-1} 
\usepackage{graphicx}
\usepackage{stmaryrd}
\usepackage{dcolumn}
\usepackage{bm}
\usepackage[usenames,dvipsnames]{xcolor}
\usepackage{multirow}
\usepackage{times}
\usepackage{chngcntr}
\usepackage{hyperref}
\usepackage{lineno} 

\definecolor{codegreen}{rgb}{0,0.6,0}
\definecolor{codegray}{rgb}{0.5,0.5,0.5}
\definecolor{codepurple}{rgb}{0.58,0,0.82}
\definecolor{backcolour}{rgb}{0.95,0.95,0.92}

\usepackage{listings}

\lstdefinestyle{mystyle}{
    backgroundcolor=\color{backcolour},   
    commentstyle=\color{codegreen},
    keywordstyle=\color{magenta},
    numberstyle=\tiny\color{codegray},
    stringstyle=\color{codepurple},
    basicstyle=\ttfamily\footnotesize,
    breakatwhitespace=false,         
    breaklines=true,                 
    captionpos=b,                    
    keepspaces=true,                 
    numbers=left,                    
    numbersep=5pt,                  
    showspaces=false,                
    showstringspaces=false,
    showtabs=false,                  
    tabsize=2
}

\renewcommand{\vr}{\ensuremath{\vc{r}}}

\newcommand{\vX}{\ensuremath{\vc{X}}}
\newcommand{\vx}{\ensuremath{\vc{x}}}

\newcommand{\Gs}{\ensuremath{\mathcal{G}}}

\newcommand{\rR}{\ensuremath{\mathcal{R}}}
\newcommand{\rH}{\ensuremath{\mathcal{H}}}
\newcommand{\lR}{\ensuremath{\bar{\mathcal{R}}}}
\newcommand{\lH}{\ensuremath{\bar{\mathcal{H}}}}

\newcommand{\up}{\ensuremath{\uparrow}}
\newcommand{\dn}{\ensuremath{\downarrow}}

\makeatletter
\newcommand{\mat}[1]{{\mathpalette\mat@{#1}}}
\newcommand{\mat@}[2]{%
  \begingroup
  \sbox\z@{$\m@th#1\underline{#2}$}%
  \dimen@=\dp\z@ \advance\dimen@ -2\mat@dimen{#1}%
  \dp\z@=\dimen@
  \sbox\z@{$\m@th\underline{\box\z@}$}%
  \box\z@
  \endgroup
}
\newcommand\mat@dimen[1]{%
  \fontdimen8
  \ifx#1\displaystyle\textfont\else
  \ifx#1\textstyle\textfont\else
  \ifx#1\scriptstyle\scriptfont\else
  \scriptscriptfont\fi\fi\fi 3
}
\makeatother

\newcommand{\vc}[1]{\ensuremath{\boldsymbol{#1}}}

\newcommand{\beq}{\begin{equation}}
\newcommand{\eeq}{\end{equation}}

\newcommand{\lbl}[1]{\label{#1}}

\renewcommand{\doi}[1]{{doi:#1}}

\usepackage[normalem]{ulem} 

\usepackage{placeins}

\begin{document}


\title{Density functional Bogoliubov-de Gennes analysis of superconducting Nb and Nb(110) surfaces}

\author{Philipp R\"u{\ss}mann} 
\email[Corresponding author: ]{p.ruessmann@fz-juelich.de}
\affiliation{Peter Gr\"unberg Institut and Institute for Advanced Simulation, 
	Forschungszentrum J\"ulich and JARA, D-52425 J\"ulich, Germany}
\author{Stefan Bl\"ugel} 
\affiliation{Peter Gr\"unberg Institut and Institute for Advanced Simulation, 
	Forschungszentrum J\"ulich and JARA, D-52425 J\"ulich, Germany}


\begin{abstract}
    We report on the implementation of the Bogoliubov-de Gennes method into the {\tt JuKKR} code [\url{https://jukkr.fz-juelich.de}], an implementation of the relativistic all-electron, full-potential Korringa-Kohn-Rostoker Green function method, which allows a material-specific description of inhomogeneous superconductors and heterostructures on the basis of density functional theory.
    We describe the formalism and report on calculations for the $s$-wave superconductor Nb, a potential component of the materials platform enabling the realization of the Majorana zero modes in the field of topological quantum computing. We compare the properties of the superconducting state both in the bulk and for (110) surfaces. We compare slab calculations for different thicknesses and comment on the importance of spin-orbit coupling, the effect of surface relaxations and the influence of a softening of phonon modes on the surface for the resulting superconducting gap. 
\end{abstract}

\maketitle


\section{Introduction}

Superconductivity is a unique manifestation of quantum mechanics on a truly macroscopic
scale and plays an important role  as an ingredient for the rapidly emerging quantum technologies.\cite{RoadmapQuantum}
An important potential application of superconductivity is the design of the topological qubit that relies on the nonabelian braiding of anions. This topological protection promises qubits with stronger robustness to
certain sources of decoherence.\cite{DasSarma2015} Such anions can be realized as zero-energy boundary states known as Majorana fermions in topological superconductors \cite{Lutchyn2018, Frolov2020, Flensberg2021} -- a material class that shows exotic spinless $p$-wave superconductivity.
To date no pure material is known that realizes this form of superconductivity, but it can be engineered in specially designed heterostructures of topological materials and $s$-wave superconductors. Therefore, these material combinations are prime candidates for the realization of Majorana zero modes and provide platforms to be explored in the growing field of materials research for topologically protected quantum computing.\cite{Schuffelgen2019}

First experimental indications of the existence of Majorana zero modes were made in form of localized states at the end of chains of magnetic Fe atoms \cite{NadjPerge2014,Kim2018} or in edge states at Co islands on top of the $s$-wave superconductor Pb.\cite{Menard2017} Other platforms where signatures of Majorana zero modes were predicted and observed are nanowires proximitized with a superconductor \cite{Alicea2011, Mourik2012} or inside Abrikosov vortices in topological superconductors.\cite{PhysRevLett.114.017001} These examples illustrate that promising physical realization of Majorana zero modes require the understanding and control of interfaces and heterostructures of superconductors with materials that have a strong spin-orbit coupling and are exposed to a magnetic field or are in contact to magnetic materials.

Real experimental interfaces are however never really perfect, thickness variations, stacking faults and single defects such as vacancies or impurities leave signatures in the superconducting interface that needs to be characterized and understood.
Material-specific simulations on the basis of \textit{ab initio} electronic structure theory
are an important ingredient in the quest for materials that realize stable Majorana zero modes.

The adaptation of density functional theory (DFT) towards the description of superconductivity was pioneered in the late 80's by Oliveira, Gross, and Kohn.\cite{Oliveira1988} This approach is based on an extension of the Kohn-Sham equation using the Bogoliubov-de Gennes (BdG) formalism which generalizes the mean-field BCS theory \cite{BCS} to describe the electron- and hole-like excitations of inhomogeneous superconductors.\cite{deGennes1966} Several works followed (e.g.\ Refs.~\onlinecite{Kohn1989, Linscheid2015, Wacker1994, AkashiArita2013, Kawamura2020, Suvasini1993, Lueders2005}) which, among others, included extensions to magnetic systems \cite{Kohn1989, Linscheid2015} or time-dependent density functional theory.\cite{Wacker1994} In our work we follow the pragmatic approximation introduced by Suvasini \textit{et al.} \cite{Suvasini1993} where superconductivity is treated as an extension to the non-superconducting solution (typically referred to as the \emph{normal state}) from standard DFT for electrons. This is done in the same spirit as introducing correlation effects of localized electrons based on a Hubbard U on-site interaction within the LDA+U method.\cite{LDAU, Tombulk}

The Korringa-Kohn-Rostoker Green's function method (KKR-GF) offers the possibility of perfect embedding of interfaces, defects, adatoms, clusters or chains of atoms, which are all important geometries for the realization of topological qubits.  With the introduction of the superconductivity we introduce a new energy scale to the DFT method. Typically the superconducting gap is in the order of a meV or less, which has to be compared to energy scales DFT methods deal with. In the normal state this are band widths in the order of $10\,\mathrm{eV}$, bandgaps in the order of $0.1$ to $5\,\mathrm{eV}$, surface states that split-off from bulk bands due to surface dipoles on scales between $0.1$ to $1\,\mathrm{eV}$ and a spin-orbit splitting of states in the order of $0.1$ to $0.3\,\mathrm{eV}$. Clearly, the small energy scale introduced by the superconducting state requires a much higher numerical resolution, which translates into higher requirements on numerical truncation parameters and much more stringent requirements on the accuracy of the shape of the charge density and potential which is feasible with the KKR-GF method.
Here we follow the work of Csire \textit{et al.} \cite{Csire2015} and implement the BdG equations in the {\tt JuKKR} code \cite{jukkr} that allows density functional theory calculations within the relativistic full-potential Korringa-Kohn-Rostoker Green's function method.\cite{Bauer2013, jukkr}


We apply our implementation of the BdG formalism to the study of superconducting Nb in the bulk and in (110) oriented surfaces.
Niobium is a well studied superconductor which is also widely used in the materials optimization towards topological quantum computing applications. It is a type-II $s$-wave superconductor with a fairly large critical temperature of $T_c\approx 9.2 \,\mathrm{K}$ and a large superconducting gap in the electronic structure of $2\Delta\approx3\,\mathrm{meV}$. 
Apart from experimental studies, \cite{Dobbs1964,MacVicar1968,Bostock1973, Hahn1998, RefRelax, Odobesko2019} several \textit{ab initio} electronic structure calculations are published that investigate the normal state properties \cite{Methfessel1992, Shein2006, RefRelax, Odobesko2019} or directly calculate superconducting properties.\cite{PhysRevB.72.024546, Tombulk, Csireheterostruc2016, PhysRevB.94.104511} This abundance of previously published data allows us to benchmark our implementation.

The paper is structured as follows. In section~\ref{sec:methods} we introduce the Bogoliubov-de Gennes method and describe its implementation into the Korringa-Kohn-Rostoker Green's function method. We then present applications of this formalism to Nb, both in the bulk as well as in surfaces of Nb(110) in section~\ref{sec:results}. We discuss the dependence of the electronic structure and the resulting superconducting properties on the thickness of the thin films which we use as structural models for the (110) surface and comment on the effect of surface phonon softening. Finally we summarize our results and conclude in section~\ref{sec:conclusion}.


\section{Methods}
\label{sec:methods}

\subsection{The Bogoliubov-de Gennes method in DFT}

The Bogoliubov-de Gennes (BdG) formalism \cite{deGennes1966} allows to describe superconductivity with a mean-field Hamiltonian which is suited to be combined with density functional theory (DFT).\cite{Oliveira1988,Suvasini1993,Csire2015} In contrast to the BCS theory of superconductivity \cite{BCS}, this allows a multi-band description and the treatment of spatially inhomogeneous superconductors and interfaces of superconductors with non-superconductors.\cite{BdGbook} In the language of DFT we can define the BdG Hamiltonian as
\begin{eqnarray}
    H_{\mathrm{BdG}}(\vx) &=& 
      \left(
      \begin{array}{cc}
          -\nabla^2-E_{\mathrm{F}}+V_{\mathrm{eff}}(\vx) & 0 \\
          0 & \nabla^2+E_{\mathrm{F}}-V_{\mathrm{eff}}^*(\vx) 
      \end{array}
      \right) \nonumber\\
      &&+
      \left(
      \begin{array}{cc}
          0 & \mathcal{D}_{\mathrm{eff}}(\vx) \\
          \mathcal{D}^*_{\mathrm{eff}}(\vx) & 0
      \end{array}
      \right)\, ,
\lbl{eq:HBdG}
\end{eqnarray}
where $\vx$ denotes a position in real space, $E_\mathrm{F}$ is the Fermi energy, $V_{\mathrm{eff}}$ is the effective single particle potential for the electrons and  $\mathcal{D}_{\mathrm{eff}}(\vx)$ is the effective pairing potential (Rydberg atomic units are used). This formulation of the pairing potential in the BdG Hamiltonian highlights the local approximation that is used for the superconducting coupling in the formalism that we adopt to the KKR-GF method.\cite{Suvasini1993}
The corresponding eigenfunctions to the eigenenergies $\varepsilon_\nu$ ($\nu$ labels the orbital and spin degrees of freedom) of this Hamiltonian are spinors $\Psi_\nu(\vx) = (u_\nu(\vx), v_\nu(\vx))^T$ where $u_\nu$ and $v_\nu$ denote the electron and hole like parts of the wave functions.
In analogy to conventional DFT calculations for the normal state, the effective potentials in \eqref{eq:HBdG} are defined as \cite{Oliveira1988,Suvasini1993}
\begin{subequations}
    \begin{align}
        V_{\mathrm{eff}}(\vx) &= V_{\mathrm{ext}}(\vx) + \int \frac{\rho(\vx')}{|\vx-\vx'|} \mathrm{d}\vx' + \frac{\delta E_\mathrm{xc}[\rho,\chi]}{\delta \rho(\vx)}, \\
        \mathcal{D}_{\mathrm{eff}}(\vx) &= \frac{\delta E_\mathrm{xc}[\rho,\chi]}{\delta \chi(\vx)}.
    \end{align}
\lbl{eq:potentials}
\end{subequations}
Obviously, the effective potentials depend on the exchange correlation energy functional, which in turn depends on the normal and anomalous densities ($\rho$, $\chi$).
In this work we use the parametrization introduced by Suvasini \textit{et al.} \cite{Suvasini1993}
\begin{equation}
    E_\mathrm{xc}[\rho,\chi] = E_\mathrm{xc}^0[\rho] - \int\chi^*(\vx)\lambda\chi(\vx)\mathrm{d}\vx
    \label{eq:Exc}
\end{equation}
with the conventional exchange-correlation functional for the normal state $E_\mathrm{xc}^0[\rho]$ (e.g.\ in the local density or generalized gradient approximation) and a semi-phenomenological coupling constant $\lambda$. 
For conventional phonon mediated superconductivity, $\lambda$ can be interpreted as the electron-phonon coupling constant that leads to the formation of Cooper pairs in the superconducting state. In this approximation the effective pairing potential reduces to $\mathcal{D}_{\mathrm{eff}}(\vx)=\lambda\chi(\vx)$. 
We stress out that the coupling parameter $\lambda$ is considered local here (as evident from \eqref{eq:Exc}), but can be space-dependent\cite{Csireheterostruc2016} (in particular atom-dependent as in the case disscused below).
Finally, the normal and anomalous densities are given by
\begin{eqnarray}
    \rho(\vx) &=& 2\sum_\nu f(\varepsilon_\nu)|u_\nu(\vx)|^2 + [1-f(\varepsilon_\nu)]|v_\nu(\vx)|^2, \\
    \chi(\vx) &=& \sum_\nu [1-2f(\varepsilon_\nu)]u_\nu(\vx)v_\nu^*(\vx)
\end{eqnarray}
where $f(\varepsilon_\nu)$ is the Fermi Dirac distribution that describes the occupation of the state $\nu$ at a given temperature.

In our work we incorporate the BdG formalism into the relativistic Korringa-Kohn-Rostoker Green's function method (KKR) which is an implementation of DFT in a multiple scattering formalism.\cite{Ebert2011} In the following we will refer to this development as ``KKR-BdG''. More details on the BdG method and its incorporation into the KKR-BdG method in general can be found in the literature.\cite{Csire2015,Tombulk} Here we focus on those aspects of the theory that are relevant to our implementation of the KKR-BdG method into the relativistic and full-potential {\tt JuKKR} code package.\cite{jukkr}

\subsection{The KKR-BdG method}

In the KKR-BdG formalism, we calculate the Green's function that is defined as the resolvent of the BdG Hamiltonian $G = (\mathcal{H}_{\mathrm{BdG}}-E)^{-1}$. In the Lehmann representation the Green's function is given as $G(\vx,\vx';E) = \sum_\nu \langle\vx'|\Psi_\nu\rangle\langle\Psi_\nu|\vx\rangle / (E-\varepsilon_\nu)$. The multiple scattering formalism of the KKR method allows to decompose the Green's function into single-site (first and second term) and multiple scattering (last term) contributions \
\begin{eqnarray}
    \mat{G}^{nn'}(\vr, \vr';E) &\equiv& \mat{G}(\vX^{n}+\vr,\vX^{n'}+\vr';E)                          \nonumber\\
     &=& -i\kappa \sum_\Lambda \bigl[ \Theta(r-r') \underline{\rR}_\Lambda^{n}(\vr;E)\underline{\lH}_\Lambda^{n}(\vr';E) \nonumber\\
     &+& \Theta(r'-r)\underline{\rH}_\Lambda^{n}(\vr;E)\underline{\lR}_\Lambda^{n}(\vr';E) \bigr] \delta_{nn'}   \nonumber\\
     &+& \sum_{\Lambda,\Lambda'}\underline{\rR}_\Lambda^{n}(\vr;E)\mat{\Gs}_{\Lambda\Lambda'}^{nn'}(E)\underline{\lR}_{\Lambda'}^{n'}(\vr';E) 
     \label{eq:KKRGF}
\end{eqnarray}
where $r=|\vr|$ is the distance to the cell center, $\Theta(r)$ is the Heaviside step function, $\kappa=\sqrt{2mE+2mE^2/c^2}/\hbar$ and $\Lambda=(\ell,m,s)$ is the combined index for angular momentum ($\ell,m$) and spin ($s$) quantum numbers.
In \eqref{eq:KKRGF} we have introduced a basis of energy dependent regular and irregular wave functions $\underline{\rR}_\Lambda$ and $\underline{\rH}_\Lambda$ (the corresponding left solutions to the radial equations, which are needed for the expansion of the Green's function if spin-orbit coupling is included \cite{Bauer2013} are indicated by the bar-notation, $\overline{\cdot}$) and structure constants $\mat{\Gs}_{\Lambda\Lambda'}$. Furthermore, atom-centered coordinates $\vc{x}=\vX^n+\vr$ ($\vX^n$ is the position vector of atom $n$ in the unit cell) as well as an expansion in real spherical harmonics around the atom centers are used.
We want to emphasize that the quantities in \eqref{eq:KKRGF} are vectors and matrices in the space characterized by $\Lambda$ as well as in the particle-hole index ($\alpha$). Single and double underscored quantities, $\underline{\phantom{\,}\!\cdot\!\phantom{\,}}$ and $\mat{\phantom{\:}\!\cdot\!\phantom{\:}}$, denote vectors and matrices in $\alpha$ and $\Lambda$, respectively. This means that in spin, particle-hole and large-small-component space the dimensions are 8$\times$8 for the Green's function, 4$\times$4 for the structural Green's function and 8$\times$1 for the radial solutions.

The single-site problem of the KKR formalism (see first and second summands in \eqref{eq:KKRGF}) breaks down to finding the site-dependent wave functions $\rR$, $\rH$, $\lR$ and $\lH$. These wave functions are the eigenfunctions to the BdG Hamiltonian of a single atom [of site $n$ with potential $V^n(\vr)$] embedded into a reference system. The reference system is traditionally the potential free space where the site-dependent wave functions are analytically known to be spherical Bessel and Hankel functions.

All wave functions are further expanded in real spherical harmonics $\mat{Y}_{\Lambda}(\hat{\vr})=\delta_{\alpha\alpha'}\delta_{ss'}Y_\ell^m(\hat{\vr})$ around the atom centers $\vX^n$ which gives the expansion for the regular right solution
\begin{equation}
\underline{\rR}_{\Lambda}^n(\vr)=\sum\limits_{\Lambda'}\,\frac{1}{r}\underline{\rR}^n_{\Lambda',\Lambda}(r) \mat{Y}_{\Lambda'}(\hat{\vr})
\end{equation}
where $\hat{\vr}=\vr/|\vr|$, and the irregular ($\rH$), regular left ($\lR$) and irregular left ($\lH$) solutions are expanded in the same way.
The coefficients of the wave functions are determined by the set of coupled radial equations (dropping the site index $n$ for simplicity)\cite{KKRbook, Bauer2013}
\begin{widetext}
\begin{equation}
    \sum_{\Lambda''} \left(\begin{array}{cc}
        (-1)^\alpha E - E_\mathrm{F} -\frac{1}{2M(r)}\frac{\ell(\ell+1)}{r^2}\delta_{\ell,\ell''}
        & 
            \frac{\mathrm{d}}{\mathrm{d}r}+\frac{1}{r} \\
        \frac{\mathrm{d}}{\mathrm{d}r}-\frac{1}{r}
        &
            -2M(r)
    \end{array}\right) \underline{\rR}_{\Lambda''\Lambda'}^{\alpha\alpha'}(r; E) = \sum_{\Lambda''\alpha''} \mat{V}_{\Lambda\Lambda''}^{\alpha\alpha''}(r)\,  \underline{\rR}_{\Lambda''\Lambda'}^{\alpha'',\alpha'}(r; E)\, ,
\lbl{eq:radial}
\end{equation}
\end{widetext}
where $M(r)$ is the relativistic mass of an electron or hole. The two components of the $\underline{\rR}$ vector presented here correspond to the large
and small relativistic components of the wave functions. In our implementation, \eqref{eq:radial} is solved using the Lippmann-Schwinger equation starting from the analytically known solutions to the free space. 
The Lippmann-Schwinger equation is solved algebraically by expanding the radial solutions in Chebychev polynomials within radial panels fully covering the atomic radius and subsequently coupling the solution in adjacent panels. The panels become gradually denser close to the origin in order to account for the singularity of the potential. \cite{Bauer2013,ZellerChebychev} The source terms for the Lippmann-Schwinger equation are the solutions to the potential in free space, which are analytically known to be spherical Bessel and Hankel functions. \cite{KKRbook,Csire2015}
In spin and electron/hole space the potential term $\mat{V}_{\Lambda\Lambda'}^{\alpha\alpha'}(r)$ in \eqref{eq:radial} can be written as [the radial dependence is implied in all terms and $L=(\ell,m)$]

\begin{widetext}
\begin{equation}
    \mat{V}_{\Lambda\Lambda'}^{\alpha\alpha'} = \left(\begin{array}{cccc}
        V_{LL'}^{\rm SRA, \up\up}+V_{LL'}^{\rm SOC,\up\up} & V_{LL'}^{\rm SOC,\up\dn}  & 0 & \delta_{LL'}\mathcal{D} \\
        V_{LL'}^{\rm SOC,\dn\up} &  V_{LL'}^{\rm SRA,\dn\dn}+V_{LL'}^{\rm SOC,\dn\dn} & -\delta_{LL'}\mathcal{D} & 0 \\
        0 & -\delta_{LL'}\mathcal{D}^* &  -\bigl(V_{LL'}^{\rm SRA,\up\up}+V_{LL'}^{\rm SOC,\up\up}\bigr)^* & -\bigl(V_{LL'}^{\rm SOC,\up\dn}\bigr)^* \\
         \delta_{LL'}\mathcal{D}^* & 0 & -\bigl(V_{LL'}^{S\rm OC,\dn\up}\bigr)^* &  -\bigl(V_{LL'}^{\rm SRA,\dn\dn}+V_{LL'}^{S\rm OC,\dn\dn}\bigr)^*
    \end{array}\right).
\lbl{eq:potential}
\end{equation}
\end{widetext}
It contains $V^{\rm SRA}$, the part of the potential subject to the scalar relativistic approximation (SRA, the SRA potential is collinear within each cell in the present implementation), the spin-orbit coupling potential $V^{\rm SOC}$ and the superconducting pairing potential $\mathcal{D}$ (that is taken as a real quantity in the present work). In this work, we have assumed $s$-wave pairing for superconductivity and apply the pairing potential only within the muffin-tin sphere around the atoms, which allows the simplification $\mat{\mathcal{D}}(r) = i\mat{\sigma}_y\delta_{LL'}\mathcal{D}(r)$  (with $\delta_{LL'}=\delta_{\ell,\ell'}\delta_{m,m'}$).
Therefore, only electrons with spin up couple to holes with spin down and vice versa. Furthermore, the matrix form of \eqref{eq:radial} neglects additional couplings between the large and small components of the wave functions coming from the potential matrix. SOC is then included on top of the SRA wave functions in the Pauli form. We find the spin-diagonal contributions, $V_{LL'}^{{\rm SOC},\up\up}$ and $V_{LL'}^{{\rm SOC},\dn\dn}$ on the diagonal elements of the potential matrix $\mat{V}_{\Lambda\Lambda'}^{\alpha\alpha'}$ and the spin-flip terms $V_{LL'}^{{\rm SOC},\up\dn}$ and $V_{LL'}^{{\rm SOC},\dn\up}$ on the off-diagonal ones in spin space. The Pauli spin-orbit coupling is an approximation that adds to the clarity of the scientific analysis of the results and allows to simplify the calculation of the radial wave functions, which speeds up the calculations significantly. This approximation has been demonstrated to agree well with the full Dirac solution of the radial equations. \cite{ZimmermannAHE, SaundersonPb2021}
In the spherical harmonics basis we use, the superconducting coupling is then given by $\mathcal{D}(r)=\mathrm{Tr}_L\left[\lambda\chi_L(r)\right]$. \cite{Csire2015} In general, we allow the coupling parameter $\lambda$ to vary from one site to the next, which allows the description of heterostructures from different superconductors and non-superconductors. \cite{Csireheterostruc2016}

From the wave functions the single-site $t$-matrices can be calculated with the expression
\begin{equation}
    \mat{t}^n(E) = \int dr\, \mat{\bar{J}}^n(r;E) \mat{V}^n(r) \mat{\rR}^n(r;E)
\lbl{eq:tmat}
\end{equation}
that contains all the scattering properties of the atom $n$ with potential $\mat{V}^n$ embedded into the reference system. For the free space as the reference system, $\bigl(\mat{\bar{J}}^n\bigr)_{\Lambda\alpha\Lambda'\alpha'} = \delta_{\alpha\alpha'}\delta_{\Lambda\Lambda'}\bar{J}_{\Lambda\alpha}^{n}$ are the spherical Bessel functions which is diagonal in the electron/hole index.\cite{Csire2015} The solution for the multiple scattering part of the Green's function  \eqref{eq:KKRGF} then demands to find the structure constants, which are given by the algebraic Dyson equation $\mat{\Gs}=\mat{\Gs}^0+\mat{\Gs}^0\,\mat{t}\,\mat{\Gs}$. Here the single-site $\mat{t}$-matrix and thus the structure constants $\mat{\Gs}$ may contain off-diagonal elements in the electron/hole sub-spaces if the potential contains terms that couple the electron and hole blocks (i.e.\ if $\mathcal{D}$ does not vanish). The reference structure constants $\mat{\Gs}^0$ are taken from the potential free case without superconducting coupling and are therefore diagonal in the electron/hole index $\alpha$.

Finally, the normal density of cell $n$ follows from the Green's function \cite{Csire2015}
\begin{eqnarray}
    \rho_n(\vr) &=& -\frac{1}{\pi} \mathrm{Im} \int\limits_{-\infty}^{\infty} \mathrm{d}E \, \mathrm{Tr}_{\Lambda} \Bigl[ f(E-E_\mathrm{F}) G_{\Lambda\Lambda'}^{nn, \rm ee}(\vr,\vr;E) \nonumber \\
                &&+ \bigl[1-f(E-E_\mathrm{F})\bigr] G_{\Lambda\Lambda'}^{nn, \rm hh}(\vr,\vr;E) \Bigr]\, ,
    \label{eq:rho}
\end{eqnarray}
where the trace $\mathrm{Tr}_{\Lambda}$ acts only on $\Lambda=(\ell,m,s)$, the ee/hh superscripts denote the electron-electron / hole-hole blocks of the Green's function and $f(E-E_\mathrm{F})$ is the Fermi Dirac distribution.
Analogously, the anomalous density
\begin{eqnarray}
    \chi_n(\vr) = -\frac{1}{\pi} \mathrm{Im} \int\limits_{-\infty}^{\infty} \mathrm{d}E \, \mathrm{Tr}_{\Lambda} &\Bigl[& \bigl[1-2f(E-E_\mathrm{F})\bigr] \nonumber\\
    && \times\, G_{\Lambda\Lambda'}^{nn, \rm eh}(\vr,\vr;E) \Bigr] \label{eq:chi}
\end{eqnarray}
follows from the off-diagonal parts of the Green's function in electron/hole space. \cite{Csire2015}
In summary, the KKR-BdG self-consistency cycle consists of (i) solving the single-site problem by finding the solution to the radial equations, (ii) solving the algebraic Dyson equation for the multiple scattering part of the Green's function and (iii) constructing the normal and anomalous densities that are finally fed back into \eqref{eq:potentials}. This procedure is iterated until the normal and anomalous densities are converged.
However, the exponential behavior of the size of the superconducting gap with the size of the electron-phonon coupling constant $\lambda$, $\Delta\sim\exp \bigl( -\frac{1}{\lambda n(E_\mathrm{F})} \bigr)$ that follows from the BCS theory \cite{BCS}, limits the range of $\lambda$ where a superconducting gap can be resolved numerically in practical calculations. This can imply that for coupling constants $\lambda$ below a critical value,  the terms that couple the electron/hole blocks in the potential may vanish during the self-consistency and the non-superconducting solution is recovered. \cite{Tombulk} This numerical inaccuracy  can be remedied by using more $k$-points in the Brillouin zone integration and more energy points in an energy contour that approaches the real axis more closely, to a point where it is practically infeasible. Nevertheless, sub-meV gap sizes that are needed in practice can be done rountinely and do not limit the applicability of this method for relevant systems. The energy integration in \eqref{eq:rho} and \eqref{eq:chi} is done in the complex plane which uses the analytical properties of the Green's function.

\subsection{Computational details}

\begin{figure}
    \centering
    \includegraphics[width=8cm]{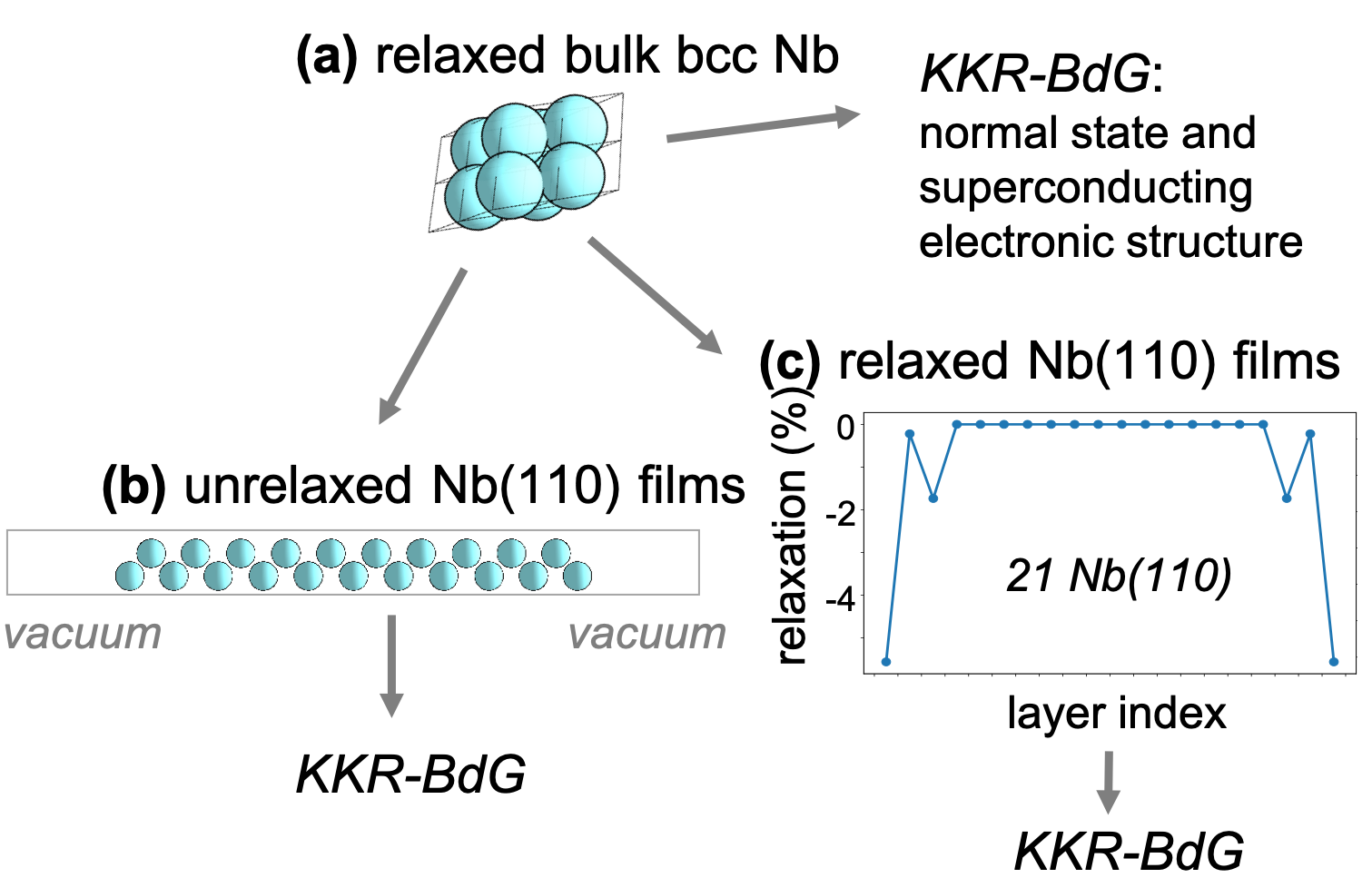}
    \caption{
    Calculation setup for bulk bcc Nb and Nb(110) surfaces. Starting from the relaxed bulk crystal structure of Nb (a) we construct thin films of Nb(110) consisting of 9, 15 and 21 layers that serve as a structural model for the Nb(110) surface. We perform calculations in the slab geometry using the layer distances of the bulk Nb crystal (b) and compare this to the relaxed geometry (c) where the first three layer distances are allowed to relax. The relaxation of the inter-layer distances shown in (c) are given relative to the inter-layer distances of the ideal bcc(110) surface. For bulk and surface geometries we calculate the electronic structure in the normal and superconducting state using the KKR-BdG method.
    }
    \label{fig:CalcOverview}
\end{figure}

In this work we apply the KKR-BdG formalism to bulk Nb and the Nb(110) surface. To describe the Nb(110) surface we use thin films of different thicknesses as structural model. We first relax the bulk lattice constant of Nb and subsequently a 13 layer thick slab of Nb(110) using the all-electron DFT code {\tt Fleur}. \cite{fleur}
We find a lattice constant of $3.32\,\mathrm{\AA}$ which is in good agreement with previous theoretical values \cite{RefRelax, Odobesko2019} as well as with the experimental value of $3.30\,\mathrm{\AA}$. \cite{RefRelax}
The relaxations in the Nb(110) thin film are carried out for the first three inter-layer distances only and the in-plane lattice constant as well as the position of the middle layers are kept fixed at the relaxed bulk values. The effect of the relaxation is summarized in Fig.~\ref{fig:CalcOverview}c. We reuse the values of the relaxed 13 layer film to model slabs of different thickness using the KKR-BdG method and study the impact of the film thickness on the superconducting state. The thin film calculations of this work are done in the slab geometry that uses periodic boundary conditions only in two directions.\cite{Wildberger1997}
The layout of the series of DFT calculations we perform for this work are summarized in Figure~\ref{fig:CalcOverview}, where in (a) the bcc crystal structure of bulk Nb and in (b) the (110) surface are shown. The amount of relaxation of the surface layers in the relaxed (110) thin films are shown in Fig.~\ref{fig:CalcOverview}(c). The most pronounced effect is the inward relaxation of the surface layer which is in agreement with previously published data for the Nb(110) surface. \cite{Methfessel1992, Shein2006, Odobesko2019}

The KKR-BdG calculations start with a self-consistent calculation of the normal state. The Fermi level is then fixed and the BdG equations are solved self-consistently starting from the converged normal state calculation. If the value of the coupling constant is large enough, \cite{Tombulk} a superconducting gap will open around the Fermi level that is typically in the order of meV. The small size of the superconducting gap also implies that the anomalous density is sharply peaked around the Fermi level (see for example Fig.~\ref{fig:adens}a), requiring much higher energy resolution close to $E_\mathrm{F}$ for the energy integration in \eqref{eq:chi} than is usual for the normal-state calculations.
Therefore, in the superconducting calculations we use a semi-circular energy contour integration in the complex plane with energy grid points getting exponentially denser when approaching the Fermi energy. The energy points that are closest to $E_\mathrm{F}$ require especially accurate calculations, which in the context of the screened KKR method \cite{Wildberger1997} is achieve with large screening cluster sizes around each atom and a much denser $k$-point sampling in the Brillouin-zone integration. \cite{Tombulk}
For the KKR-BdG calculations, in this work we use an angular momentum cutoff of $\ell_{max}=3$, 
we include corrections for the exact shape of the cells around the atoms (i.e.\ full-potential calculations) \cite{Stefanou1990,Stefanou1991} and our semi-circle energy contour consist of 48 energy points. In the final KKR-BdG self-consistency cycle, a screening cluster radius of $10.05\,\mathrm{\AA}$ ($11.61\,\mathrm{\AA}$) is used,  which includes 259 (339) sites and a dense $k$-mesh for the Brillouin zone integration of $100^3$ ($100^2$) for the bulk (surface) calculations. The final density of states calculations with sub-meV energy resolution are are carried out for a smearing temperature of $0.1\,\mathrm{K}$ with a $k$-mesh of $400^3$ ($1000^2$) for the bulk (surface) calculations.

All our DFT calculations use the GGA-PBE exchange correlation functional \cite{PBE} for the normal state. The series of calculations in this study are orchestrated using the \emph{AiiDA-Fleur} \cite{aiida-fleur, aiida-fleur2} (used for the relaxations with the Fleur code) and \emph{AiiDA-KKR} \cite{aiida-kkr} (used for the KKR-BdG calculations) plugins to the AiiDA infrastructure. Within AiiDA, \cite{aiida1,aiida2} calculations are submitted to supercomputers while keeping track of all inputs/outputs and storing of the results of the calculations to a database. This keeps track of the provenance between calculations represented by a directed acyclic graph in the database which ensures reproducability of our results according to the FAIR principles of {findable}, {accessible}, {interoperable}, and {reusable} data. \cite{Wilkinson2016} The dataset of this work including the full provenance of our calculations is made publicly available in the materials cloud repository. \cite{Talirz2020,doi-dataset} We point out that this dataset also contains a short tutorial for a simple example of superconducting bulk Nb using the KKR-BdG method done with AiiDA-KKR.

\section{Results and Discussion}
\label{sec:results}

\subsection{Bulk Nb}

\begin{figure}
    \centering
    \includegraphics[width=8cm]{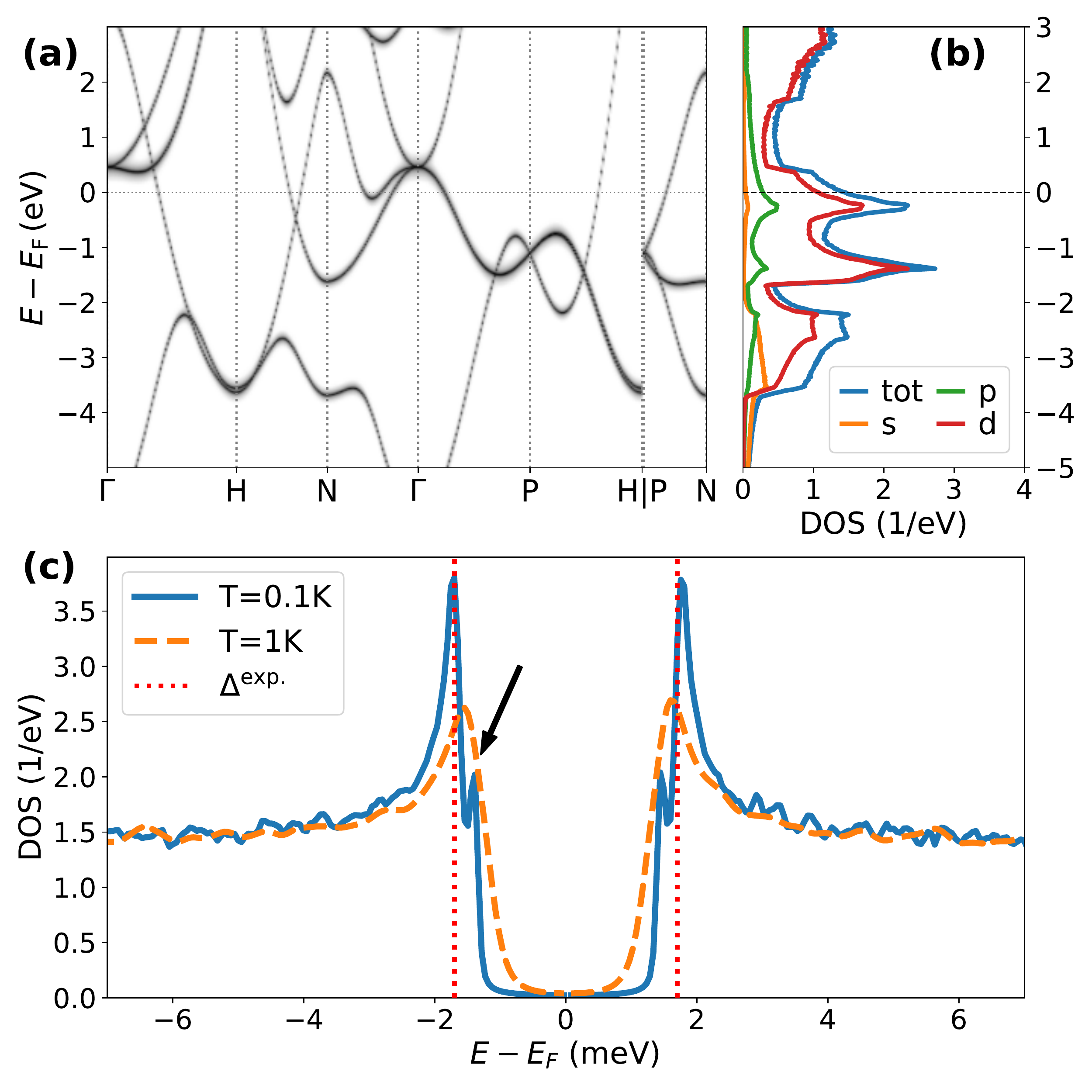}
    \caption{Normal state and superconducting electronic structure of bulk Nb. For the normal state the band structure (a) is shown
    together with the corresponding density of states (b). The DOS in the superconducting state (c) is shown for the electron component only and it reveals the opening of a gap around $E_\mathrm{F}$. For a high enough energy resolution (i.e. low enough smearing temperatures $T$) a gap anisotropy at $\pm1.4\,\mathrm{meV}$  can be resolved (highlighted by the black arrow). The coupling parameter $\lambda$ was chosen such that the superconducting gap matches the experimental value ($\Delta^{\mathrm{exp.}}$ indicates position of the experimental coherence peaks \cite{Odobesko2019}). 
    }
    \label{fig:bulk}
\end{figure}

We start our discussion with the electronic structure of bulk Nb. The normal state's electronic structure is presented in Fig.~\ref{fig:bulk}(a,b) where the band structure is shown
alongside the density of states (the Brillouin zone of the bcc crystal, where the high symmetry points are indicated can be found in Fig.~\ref{fig:SOC}). 
The superconducting state is then simulated using a coupling constant of $\lambda=1.11\,\mathrm{eV}$ which was found to reproduce the size of the superconducting gap that was measured in a recent scanning tunneling microscopy (STM) experiment on the clean Nb(110) surface.\cite{Odobesko2019} This value is in good agreement with earlier results by Saunderson \textit{et al.} who used $\lambda = 1.17\mathrm{eV}$ for calculations based on the local density approximation (LDA).\cite{Tombulk} The corresponding density of states is shown in a small energy window around the Fermi level in Fig.~\ref{fig:bulk}(c). A close inspection of the superconducting DOS reveals a shoulder at the edge of the superconducting gap (highlighted by the black arrow). This is a signature of the small gap anisotropy in Nb which agrees well with the prediction made in earlier calculations \cite{Tombulk} as well as with experimental observations.\cite{Dobbs1964,MacVicar1968,Bostock1973,Hahn1998} It is worthwhile noting that an increased temperature smearing (shown as the orange dashed line) of $1\,\mathrm{K}$ strongly suppresses the signature of the gap anisotropy in the DOS which may explain why no gap anisotropy is visible in the STM experiments of Ref.~\cite{Odobesko2019} that were done at $1.3\,\mathrm{K}$.

\begin{figure}
    \centering
    \includegraphics[width=8cm]{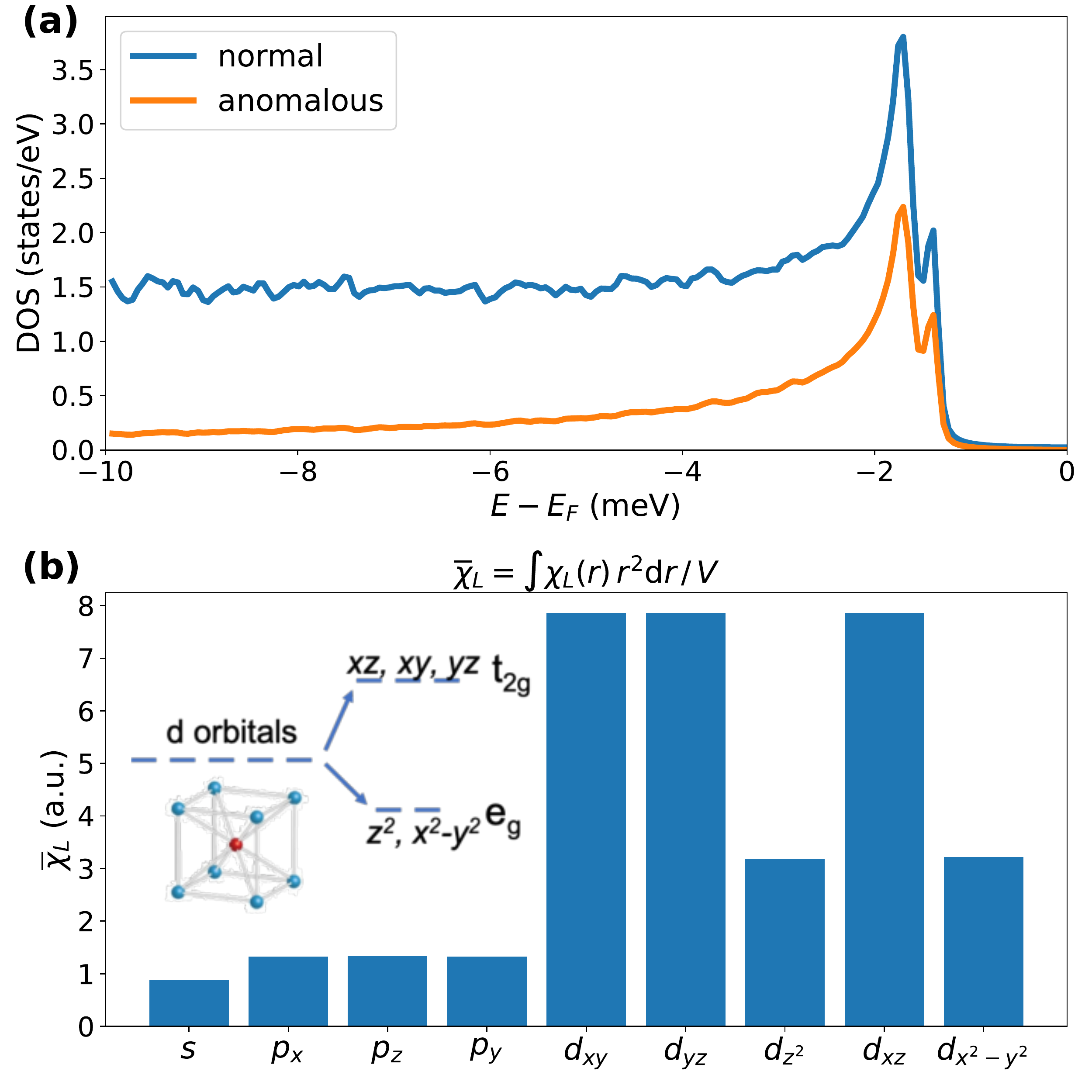}
    \caption{Properties of the superconducting order parameter. (a) Energy dependence of the anomalous density (orange) together with the electronic density of states (blue, labelled `normal') of superconducting bulk Niobium. The radially averaged anomalous density (b) reflects the orbital structure and the crystal field splitting (shown as an inset) which is inherited from the normal state electronic structure of bulk Nb.}
    \label{fig:adens}
\end{figure}

The description of superconductivity within the KKR-BdG method relies on the calculation of the anomalous density (cf.\ \eqref{eq:chi}). We will now discuss some properties of the anomalous density, which is the superconducting order parameter in the KKR-BdG formalism, and its connection to the normal state electronic structure of bulk Nb. Figure~\ref{fig:adens}a shows the energy dependence of the anomalous density and the electronic density in terms of their respective density of states for superconducting bulk Niobium. On the $\mathrm{meV}$ energy scale the normal state DOS shows almost no variation and would be a flat line around $1.5\, \mathrm{eV}^{-1}$. The opening of the superconducting gap results in a coherence peak around $E-E_{\mathrm{F}} = -2\,\mathrm{meV}$ with the gap anisotropy that is discussed in Fig.~\ref{fig:bulk}. The coherence peak can be traced back to the anomalous density which shows the same features of the coherence peaks including gap anisotropy and the apearance of the superconducting gap. It is noteworthy that the anomalous density quickly decays in energy and has almost no contribution $10 \,\mathrm{meV}$ away from the coherence peaks. This highlights the extreme accuracy around the Fermi level that is needed in the KKR-BdG calculations.
The energy integrated and radially averaged anomalous density is shown in Fig.~\ref{fig:adens}b including its decomposition into the orbital character. We see that the anomalous density mainly has contributions from the $d$ orbitals that dominate the electronic structure of Nb at the Fermi level. The crystal field of the bcc crystal leads to a splitting of the $d$ orbitals into lower lying $e_g$ and higher lying $t_{2g}$ states. This orbital character is recovered in the orbital decomposition of the anomalous density that is inherited from the normal state electronic structure of bulk Nb.
These results highlight the importance of the normal state's electronic structure around the Fermi level for the properties of the superconductivity and shows that the KKR-BdG method is capable of dealing with multi-band superconductors as it was pointed out by Saunderson \textit{et al}.\cite{Tombulk}

We compared calculation with and without spin-orbit coupling and found no significant difference in the superconducting properties as discussed in Appendix~\ref{app:SOC}. Thus for the discussion in the main text we work in the scalar relativistic approximation and neglect effects of SOC in the remainder.


\subsection{Nb(110) surfaces}

\begin{figure*}
    \centering
    \includegraphics[width=1.0\linewidth]{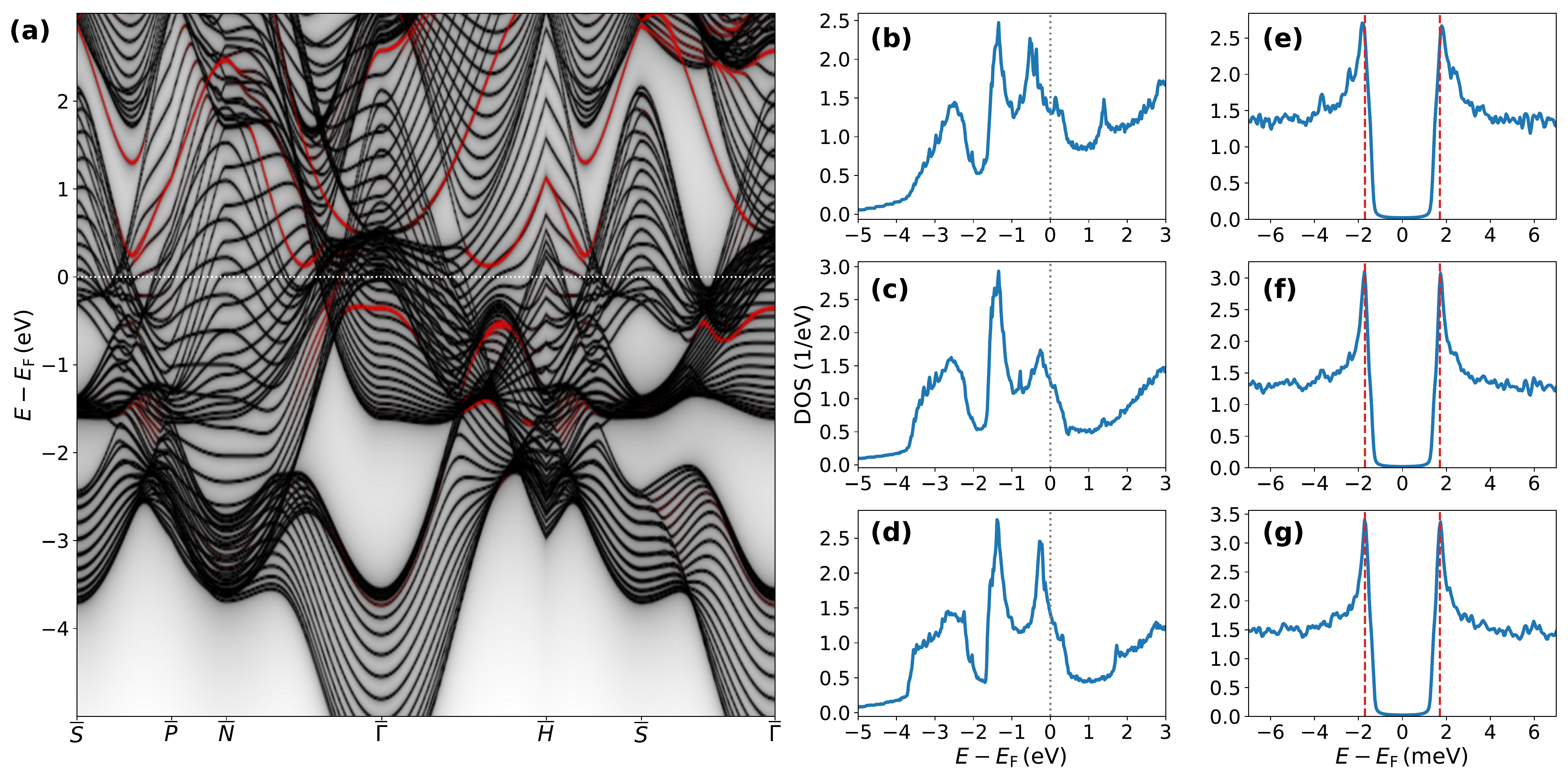}
    \caption{
    Electronic band structure
    and local density of states (DOS) in the normal state (a,b-d) and in the superconducting state around the Fermi level (e-g) for the unrelaxed 21 layer Nb(110) thin film. The black-red color scale in (a) indicates the surface localization of the wave functions (red: localized in the surface layer, black: localized in the rest of the film). The normal state DOS (b-d) and the superconducting DOS (e-g) are shown (from top to bottom) for the first two surface layers and the central Nb layer in the thin film. The vertical dashed lines in (e-g) are guides to the eye that highlight the layer independence of the superconducting gap throughout the Nb(110) thin film. (Note the difference in energy scale for the displayed DOS.) }
    \label{fig:surfaceDOS}
\end{figure*}

We now turn our attention to the KKR-BdG calculation for Nb(110) surfaces. 
We start with a discussion of the electronic structure of an unrelaxed 21 layer Nb(110) thin film that is shown in Fig.~\ref{fig:surfaceDOS}. 
Figure~\ref{fig:surfaceDOS}(a) shows the band structure for this thin film in the normal state along paths connecting high-symmetry points of the surface Brillouin zone (SBZ, see Fig.~\ref{fig:SOC} in the Appendix).
The presence of the surface leads to a projection of the bulk bands into the 2D SBZ and the finite size of the thin film leads to quantum well states which are observed as discrete lines in the band structure.
The color code in Fig.~\ref{fig:surfaceDOS}(a) indicates the localization of the wave function throughout the thin film. This is shown in terms of the ratio of the spectral weight in the surface layer compared to the sum over all layers (red bands are localized to $\ge 80\%$ on the surface). We can see that the contributions of the surface layers to the bands that cross the Fermi energy is not significant. Therefore no strong change in the superconducting properties due to surface state is to be expected.
A large number of bands cross the Fermi level which is why for the discussion of the superconducting properties we turn to the density of states integrated over the Brillouin zone that is shown layer-resolved in Fig.~\ref{fig:surfaceDOS}(b-g) for the normal (b-d) and superconducting state(e-g), respectively.

Lowering the coordination number due to the presence of a surface visibly changes the local electronic structure. Electrons from the $s$ and $p$ derived bands (mainly around $E-E_\mathrm{F} = -3\,\mathrm{eV}$) leak into the vacuum and deplete the local DOS of the surface layer and the lower symmetry leads to splittings in the $d$ electron derived bands which is clearly visible around $E-E_\mathrm{F}=-0.5\,\mathrm{eV}$.
Here the superconducting state is modelled using a uniform coupling constant $\lambda$ throughout the thin film and it is scaled to reproduce the experimentally observed size of the gap. We can see that the resulting superconducting gap is largely layer independent and has the same size throughout the full 21 layer film. We have performed this analysis also for thinner films (9 and 15 layers) and compared the superconducting electronic structure for both relaxed and unrelaxed geometries and found always that the resulting superconducting gap shows no layer dependence (not shown here). This result may be counter intuitive as the density of states at the Fermi level does show changes with the number of layers and the relaxations that are taken into account.
In order to reproduce the experimental size of the superconducting gap we used different scaling factors for the coupling constant $\lambda$ for the different thin film thicknesses and relaxations ranging from $\lambda=1.11\,\mathrm{eV}$ in the bulk calculation to $\lambda=1.23\,\mathrm{eV}$ for the relaxed surface.
We can understand this from the fact that the superconducting coherence length of Niobium is much larger than the inter-layer distance in the Nb(110) thin films and also larger than the largest film thickness we considered. 
This is expected since the superconducting gap in the density of states in Nb(110) results from an average of the anomalous densities over distances of the length scale of the coherence length.

\begin{figure}
    \centering
    \includegraphics[width=8cm]{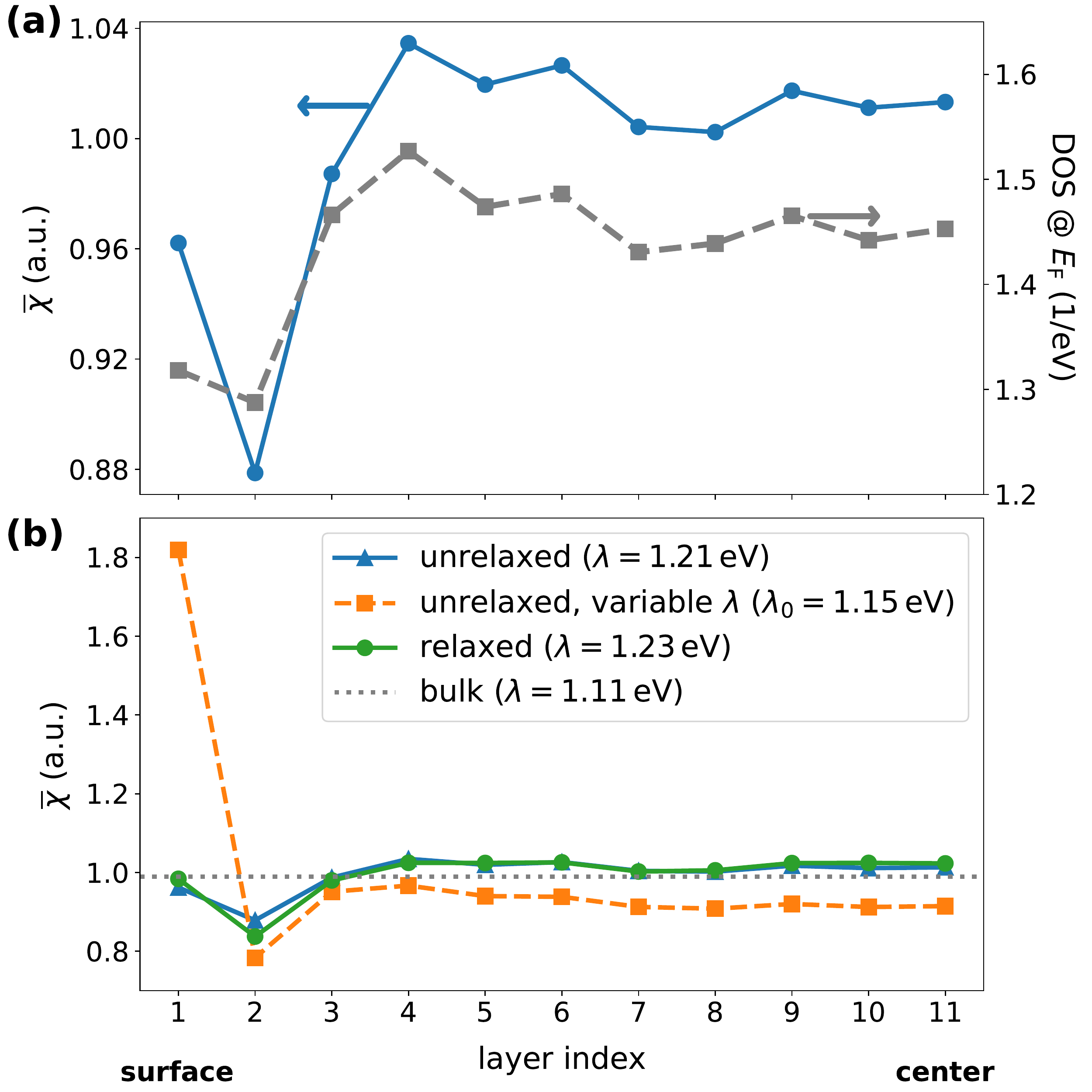}
    \caption{(a) Layer dependence of the anomalous density and the density of states at the Fermi level of an unrelaxed 21 layer Nb(110) thin film with uniform coupling constant $\lambda$ throughout the thin film. The anomalous density is shown in terms of the radial average over the muffin-tin sphere around the atoms $\overline{\chi} = \int_{V_{\mathrm{MT}}} \mathrm{Tr}_\Lambda \bigl[ \chi_{\Lambda}(r) \bigr] \, r^2\mathrm{d}r / V_{\mathrm{MT}}$. (b) Layer resolved average anomalous density for 21 layer thick films with (i) constant $\lambda$ without structural relaxations, (ii) for the unrelaxed film with varying $\lambda$ that models surface enhancement of the electron-phonon coupling and (iii) for the relaxed film with constant $\lambda$ throughout the layers.
    }
    \label{fig:adensLayer} 
\end{figure}

We can substantiate our claim of the layer independence of the superconducting gap with an analysis of the layer-dependence of the anomalous density that is shown for the 21 layer unrelaxed Nb(110) film in Fig.~\ref{fig:adensLayer}a together with the layer resolved density of states of the normal state at the Fermi energy. We can see that the average anomalous density varies by roughly 15\% throughout the thin film and follows the local variation of the electronic structure as seen in the normal states DOS at the Fermi level. We stress that these variations come purely from the changes in the electronic structure as we have chosen our effective coupling constant $\lambda$ to be constant throughout the film. The values of the coupling constant $\lambda$ for the different film thicknesses and relaxations are scaled individually to reproduce the same superconducting gap size in the density of states as in the bulk. Figure~\ref{fig:adensLayer}b shows a comparison of the resulting layer resolved anomalous densities for the 21 layer thick Nb(110) with and without relaxed inter layer distances of the 3 outermost layers. The changes to the electronic structure only lead to small changes of the density of states at $E_\mathrm{F}$ and therefore only a small renormalization of the effective coupling constant $\lambda$.

Finally we considered a spatial variation of the local electron-phonon coupling constant which is included in our formalism in the size of the effective coupling constant $\lambda$. The connection of the normal state electronic structure at the Fermi level and the electron-phonon interaction is captured in the McMillan formula \cite{McMillan1968}
\begin{equation}
    \lambda_i = \frac{D_i(E_\mathrm{F})\langle g^2_i\rangle}{M_i\langle\omega^2_i\rangle}
\end{equation}
where $i$ is the layer index, $D_i(E_\mathrm{F})\langle g^2_i\rangle$ is the McMillan-Hopfield parameter that contains the density of states at the Fermi energy in layer $i$ and its average electron-phonon matrix element, $M_i$ is the mass of atom $i$, and $\langle\omega^2_i\rangle$ is the average squared phonon frequency. The presence of the surface in Nb(110) introduces a softening of the surface phonons which results in a smaller average phonon frequency and consequently in an increased electron-phonon coupling constant that is roughly 50\% larger in the surface layer.\cite{CsireSchoenecker2016} We use this increase in the surface electron-phonon coupling constant 
($\lambda=1.15\,\mathrm{eV}\equiv\lambda_0$ throughout the thin film except for $\lambda=1.5\lambda_0$ in the surface layers) 
and perform KKR-BdG calculations for the unrelaxed 21 layer thin film. The resulting layer resolved average anomalous density is shown in Fig.~\ref{fig:adensLayer}b (dashed orange line). The strong increase in the anomalous density of the surface layer can be attributed to the 50\% increase in the coupling constant.
Although the anomalous density shows such a strong modulation throughout the layers, the superconducting gap is found to be uniform over the thickness of the thin film. In order to reproduce the gap size of the bulk we had to rescale the $\lambda$ values of all layers (i.e.\ $\lambda_0=1.15\,\mathrm{eV}<\lambda=1.21\,\mathrm{eV}$ for the unrelaxed film with constant coupling constant). The smaller contribution of the layers in the center of the film now counteract the increase of the electron-phonon coupling constant in the surface layers and their therefore strong increased anomalous density. 

In all our calculations (i.e.\ with and without relaxations, constant or varying $\lambda$ throughout the layers of the thin film) we calibrated the coupling constants such that the resulting superconducting gap in the density of states reproduces the experimentally observed value 
from STM experiments \cite{Odobesko2019} that we also used to calibrate the coupling constant for the bulk Nb calculations.
We find that the proper scaling results in an anomalous density that, if averaged over the film thickness, results in the same average anomalous density as in the bulk calculation (grey dotted line in Fig.~\ref{fig:adensLayer}b).
This result supports our observation that the superconducting gap is averaged through the layers over the length scale of the superconducting coherence length.

\section{Summary and conclusions}
\label{sec:conclusion}

In summary, we introduce our implementation of the Bogoliubov-de Gennes method into the full-potential relativistic Korringa-Kohn-Rostoker Green function code {\tt JuKKR}. We apply this to the study of the electronic structure of the $s$-wave superconductor Nb. We compare the normal and superconducting state of Nb both in the bulk and for the Nb(110) surface. As a structural model we use thin films of different thicknesses and we take into account the effect of relaxations of the surface layers. We comment on the implications of a surface phonon softening which both influence the superconducting order parameter (i.e.\ the anomalous density) but which is averaged out over the full film thickness due to the long superconducting coherence length of Nb. This results on a layer independent superconducting gap throughout slabs of different thickness.

Our approach can be generalized to the description of impurities embedded into superconductors \cite{Tomimp} and superconducting heterostructures where an accurate description of the atomic cells with the full-potential treatment is expected to be particularly important. Our demonstrated features of combining the accurate electronic structure, that can include spin-orbit coupling, and the Bogoliubov-de Gennes formalism that captures the behavior of an $s$-wave superconductor will allow us to dive into the optimization problem behind material platforms that allow the realization of Majorana fermions in solid state systems in the future. The interface to the AiiDA framework that allows automation of these calculations will there be especially beneficial as it allows to tackle the problem of a large combinatorial space with high-throughput calculations.\cite{aiida1,aiida2,aiida-kkr}








\section*{Acknowledgements}

We would like to thank Phivos Mavropoulos, Tom G.\ Saunderson and Martin Gradhand for fruitful discussions about the KKR-BdG method and its implementation. We thank Vasily Tseplyaev for his support in setting up the AiiDA-Fleur calculations used for the relaxations.
We acknowledge support by the Joint Lab Virtual Materials Design (JL-VMD) and we are grateful for computing time granted by the JARA Vergabegremium and provided on the JARA Partition part of the supercomputer CLAIX at RWTH Aachen University.
This work was funded by the Deutsche Forschungsgemeinschaft (DFG, German Research Foundation) under Germany's Excellence Strategy – Cluster of Excellence Matter and Light for Quantum Computing (ML4Q) EXC 2004/1 – 390534769.








\appendix
\counterwithin{figure}{section}

\section{Effect of spin-orbit coupling on the electronic structure of Nb and Nb(110) surfaces}
\label{app:SOC}

\begin{figure*}
    \centering
    \includegraphics[width=1.0\linewidth]{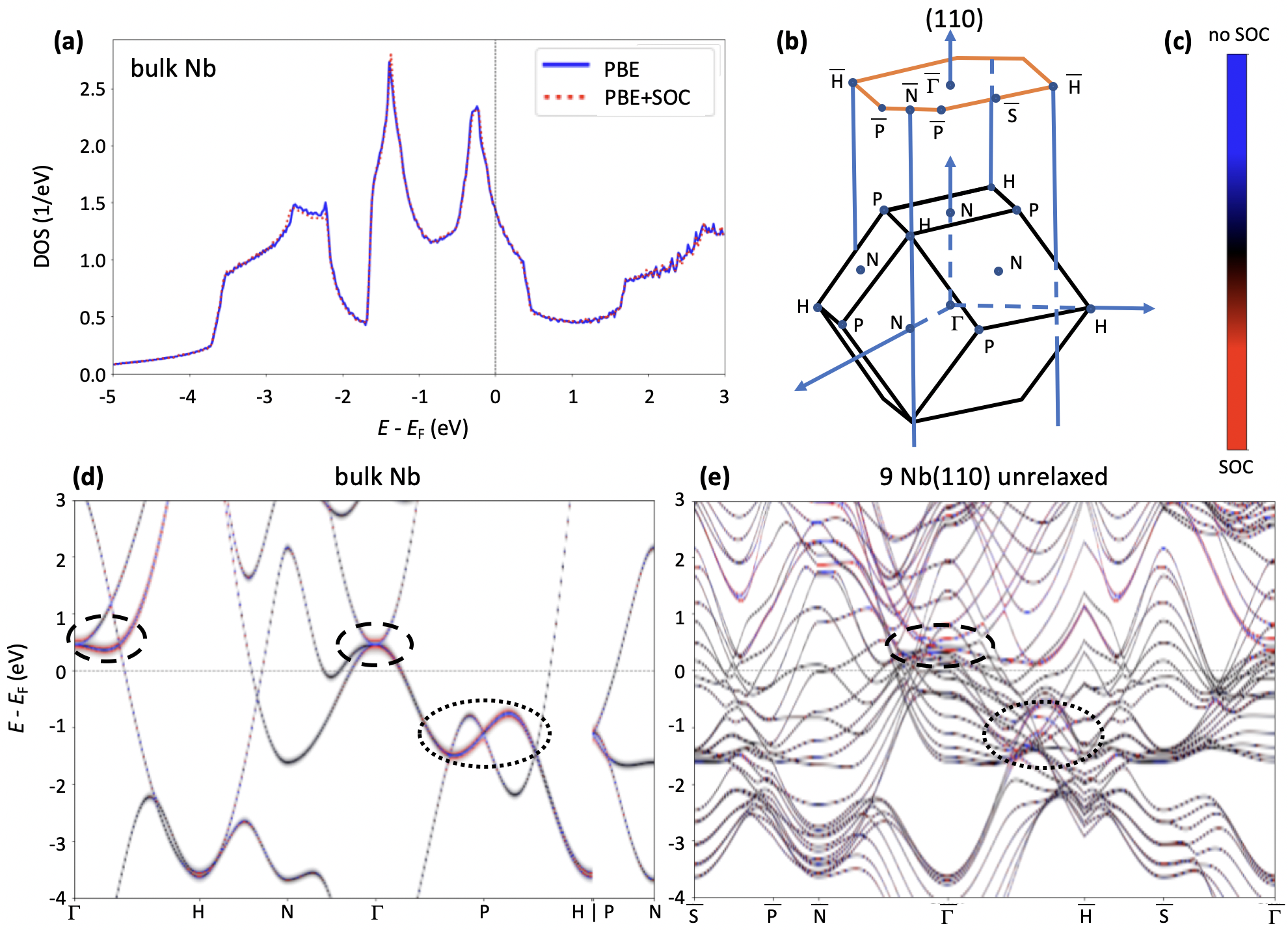} 
    \caption{Effect of spin-orbit coupling (SOC) on the normal state electronic structure in bulk Nb and on the 110 surface. (a)  Comparison of the density of states for bulk Nb without and with SOC. The change in the electronic structure with SOC for bulk Nb (d) and a Nb(110) surface (e) are shown  in terms of the difference of the Bloch spectral function corresponding to the colour scale shown in (c). The corresponding high symmetry points of the bulk and surface Brillouin zones are indicated in (b). For bulk Nb the largest changes due to SOC occur around $\Gamma$ and P and are highlighted with dashed and dotted ellipses in (d).}
    \label{fig:SOC}
\end{figure*}

In our KKR-BdG calculations for the superconducting state of bulk Nb and Nb(110) surfaces we neglect the effect of spin-orbit coupling (SOC). To justify this we first demonstrate that the normal state electronic structure of Nb around the Fermi energy does not change with SOC which then does not affect superconducting properties like the average gap. Figure~\ref{fig:SOC} compares our KKR calculations for the normal state of Nb in the scalar relativistic approximation with the corresponding calculation where SOC effects are included self-consistently. The density of states of bulk Nb that is shown in Fig.~\ref{fig:SOC}(a) reveals almost no visible change from the scalar relativistic (PBE) calculation to the calculation that includes SOC (PBE+SOC). We highlight the difference in the electronic structure of bulk Nb upon including SOC effects by taking the difference of the Bloch spectral function \cite{Ebert2011} (which visualizes the band structure) of the two calculations in Fig.~\ref{fig:SOC}(b). The red-black-blue color code indicates that SOC lifts some degeneracies in the electronic structure which leads to the formation of avoided crossings. The changes closest to $E_\mathrm{F}$ are highlighted with dashed and dotted ellipses. We want to point out that these visible changes occur not at the Fermi level but at least $\sim 0.5\,\mathrm{eV}$ away from $E_\mathrm{F}$ and therefore only leads to minimal changes of the electronic structure at $E_\mathrm{F}$.
We can draw the same conclusion for the Nb(110) surfaces. The SOC effect on the band structure is shown exemplary for the unrelaxed 9 layer film in Fig.~\ref{fig:SOC}(e). We again see that the bands that cross the Fermi level are colored in black which means that there is a negligible change in the electronic structure at $E_\mathrm{F}$ due to SOC.

To verify that the SOC effect can be neglected for the superconducting calculations of Nb we performed KKR-BdG calculations on top of the PBE+SOC calculations for both bulk Nb and the unrelaxed 9 layer Nb(110) surface. We compare these calculation to the ones presented in the main part of the manuscript. For this comparison we fixed the value of the coupling constant $\lambda$ and compared the average anomalous density $\overline{\chi}=\int_{MT} \mathrm{Tr}_L\left[\chi_L(r)\right]r^2\mathrm{d}r/V_{MT}$ which is the radial average over the muffin tin sphere. Note that this directly relates to the average size of the superconducting coupling of the BdG Hamiltonian (i.e\ $D_\mathrm{eff}$ in \eqref{eq:HBdG}).
We find that including SOC changes $\overline{\chi}$ by less than 1\% both in the bulk and in the surface calculations. Therefore we conclude that SOC effects can safely be neglected for our study of bulk Nb and Nb(110) surfaces but can be vital for the correct descriptions of other materials.

\bibliography{main}


\end{document}